# Neighborhood-based Bridge Node Centrality Tuple for Preferential Vaccination of Nodes

Natarajan Meghanathan
Jackson State University, Jackson, MS 39217, USA
Author Email: natarajan.meghanathan@jsums.edu

**Abstract**
We investigate the use of a recently proposed centrality tuple called the Neighborhood-based Bridge Node Centrality (NBNC) tuple to choose nodes for preferential vaccination so that such vaccinated nodes could provide herd immunity and reduce the spreading rate of infections in a complex real-world network. The NBNC tuple ranks nodes on the basis of the extent they play the role of bridge nodes in a network. A node is a bridge node, if when removed its neighbors are either disconnected or at least sparsely connected. We hypothesize that preferentially vaccinating such bridge nodes would block an infection to spread from a neighbor of the bridge node to an another neighbor that are otherwise not reachable to each other. We evaluate the effectiveness of using NBNC to reduce the spread of infections by conducting simulations of the spread of infections per the SIS (Susceptible-Infected-Susceptible) model on a collection of 10 complex real-world social networks. We observe the average fraction of infected nodes per round of the SIS simulations based on NBNC for preferential vaccination to be lower than that of the degree centrality-based preferential vaccination.

## 1 Introduction

With the COVID-19 pandemic creating havoc worldwide for the last few years, and the necessity of vaccinating people with boosters to protect them from getting infected with variants of the virus, it becomes imperative to identify effective strategies to preferentially vaccinate people in a social network or a community so that such vaccinated people could provide herd immunity (i.e., block the infection from spreading to the non-vaccinated people) and thereby the average number of infected people per network or community is eventually reduced. Network Science provides solution to the above problem through the notion of "Centrality" metrics. Centrality metrics quantify the topological importance of a node in the network [14]. There exists a plethora of centrality metrics, among which the degree centrality (DEG) is the most computationally-light metric that has been often used (e.g., [15]; most recent use reported in [1]) to rank the nodes for preferential vaccination under the premise that nodes with several neighbors are more likely to spread the infection from one neighbor to another neighbor and vaccinating high DEG nodes could reduce the overall number of infected nodes.

In a recent work [11], the author proposed the notion of neighborhood-based bridge node centrality (NBNC) tuple to quantify and ranked based on the extent to which nodes play the role of bridge nodes. A node is a bridge node [11, 16] if when removed the neighbors of the node either get disconnected or are more likely to be sparsely connected (i.e., have to reach through a longer path of length much greater than 2). Our hypothesis in this paper is that bridge nodes (rather than high DEG nodes) would be a better choice for preferentially vaccinating nodes to attain herd immunity and reducing the overall number of infected nodes. Our hypothesis stems from the criteria used to rank the nodes as bridge nodes per the NBNC tuple.

The NBNC tuple of a node has three entries (the number of components in the neighborhood graph of the node, the algebraic connectivity ratio of the neighborhood graph of the node, the number of vertices in the neighborhood graph, which is also the degree of the node), A node with a higher degree may still have its neighbors connected (either directly or through a multi-hop path) if the neighbor node is removed from the network. On the other hand, if a node has several components in its neighborhood graph, it is more likely that any two neighbors of the node are disconnected or connected through a long multi-hop path when the node is removed from the network. Note that DEG is also the last of the three entries in the tuples for NBNC; if two nodes cannot be differentiated based on the first two entries in the NBNC tuple, then the degree of the nodes could be used to break the tie. Hence, NBNC (whose tuple formulation includes DEG as the last entry) is a more comprehensive centrality tuple (compared to a scalar DEG centrality metric) to rank the bridge nodes for preferential vaccination.

We use the SIS (susceptible-infected-susceptible) model [17], one of the widely used models for simulating the spread of infections. The $R_0$ (basic reproduction number) for a disease [17] is defined as the number of infections an infected individual could cause to a completely susceptible population. If $R_0$ for a disease is greater than 1, then one infected individual could lead to more than one newly infected individual and the disease would keep spreading. On the other hand, if $R_0$ for a disease gets less than 1, the chances of one infected individual leading to another infected individual gets lower and the disease will eventually die down. One way to reduce the $R_0$ for a disease is to preferentially vaccinate some nodes so that even if these nodes are exposed to the infected individuals, they would not get infected; as a result, if susceptible (non-vaccinated) individuals are exposed to the vaccinated individuals (but not to the infected individuals), the infection would not spread. At the same time, the infected individuals would also eventually recover and the disease would eventually die down. Hence, the motivation for this paper is to explore the use of NBNC for preferentially vaccinating nodes in the presence of a disease spread simulation (in rounds) per the SIS model and evaluate whether this leads to a lower value for the average number of infected nodes per round of the simulation (compared to the strategy of vaccinating nodes based on node degree, the currently preferred strategy).

The rest of the paper is organized as follows: Section 2 presents the notion of the NBNC tuple, its calculation and the ranking procedure as well as illustrates all of these with a toy example graph. Section 3 presents a simulation procedure for running the SIS model on a graph in the presence of a certain fraction of vaccinated nodes per the NBNC tuple vs. DEG centrality. Section 4 presents the results of the SIS simulations conducted for a suite of 10 complex real-world networks and compares the average number of infected individuals per round incurred with the NBNC tuple vs. DEG centrality-based vaccinations. Section 5 presents related work in the literature. Section 6 concludes the paper and presents plans for future work. Throughout the paper, the terms 'node' and 'vertex', 'link' and 'edge', 'network' and 'graph' are used interchangeably. They mean the same.

## 2 Neighborhood-based Bridge Node Centrality (NBNC) Tuple

The NBNC tuple of a node $v$ is determined based on the neighborhood graph (*NG*) of the node. The neighborhood graph (*NG*) of a node $v$ comprises of just the neighbor nodes of $v$ and the edges connecting these neighbor nodes. Note that the *NG* of a node $v$ does not include $v$ and the edges incident on $v$. The NBNC tuple for a node $v$ has three entries, represented in this order: $NBNC(v) = [NG(v)^{\#comp}, NG(v)^{ACR}, |NG(v)|]$. $NG(v)^{\#comp}$ is the number of components in the neighborhood graph of node $v$. If the neighbors of node $v$ are connected (reachable to each other) even after node $v$ is removed from the network, then $NG(v)^{\#comp} = 1$; otherwise $NG(v)^{\#comp} > 1$. The second entry $NG(v)^{ACR}$ is the algebraic connectivity ratio of the neighborhood graph; the algebraic connectivity [18] of the neighborhood graph is the second Eigenvalue of the Laplacian matrix [19] of the neighborhood graph. If $NG(v)^{\#comp} > 1$, then the neighborhood graph is not connected and $NG(v)^{ACR}$ is 0; otherwise, $NG(v)^{ACR}$ is computed by dividing the algebraic connectivity by the number of nodes in the neighborhood graph (which also corresponds to the degree of node $v$). The third entry is the number of vertices in the neighborhood graph of the node, which is the degree of the vertex itself.

The following ranking criteria is used to rank nodes on the basis of the extent they play the role of bridge nodes. In summary, nodes with more components in their neighborhood graph are ranked higher. If two nodes have the same number of components in their neighborhood graph, then the tie is broken in favor of the node with sparsely connected neighborhood. If two nodes cannot be differentiated based on the first two entries, then the tie is broken in favor of larger node degree. If two nodes cannot be differentiated based on all the three entries of their NBNC tuples, then they are equally ranked. In summary, for two nodes $u$ and $v$:
(i) If $NG(u)^{\#comp} > NG(v)^{\#comp}$, then node $u$ is ranked higher.
(ii) If $NG(u)^{\#comp} = NG(v)^{\#comp}$ and $NG(u)^{ACR} < NG(v)^{ACR}$, then node $u$ is ranked higher.

(iii) If $NG(u)^{\#comp} = NG(v)^{\#comp}$ and $NG(u)^{ACR} = NG(v)^{ACR}$ and $|NG(u)| > |NG(v)|$, then node $u$ is ranked higher.

(iv) If $NG(u)^{\#comp} = NG(v)^{\#comp}$ and $NG(u)^{ACR} = NG(v)^{ACR}$ and $|NG(u)| = |NG(v)|$, then both the nodes are equally ranked.

Figure 1 presents a toy example graph that will be used in this section as well as in the next section. Figure 1 also presents the neighborhood graphs for some of the vertices of the graph; their Laplacian matrix and the Eigenvalues of the Laplacian matrix. The entries in a Laplacian matrix for a neighborhood graph are as follows: The diagonals indicate the number of neighbors for the nodes in the neighborhood graph; an entry $(u, v)$ is -1 if there is an edge between $u$ and $v$ in the neighborhood graph; otherwise, the entry is a 0.

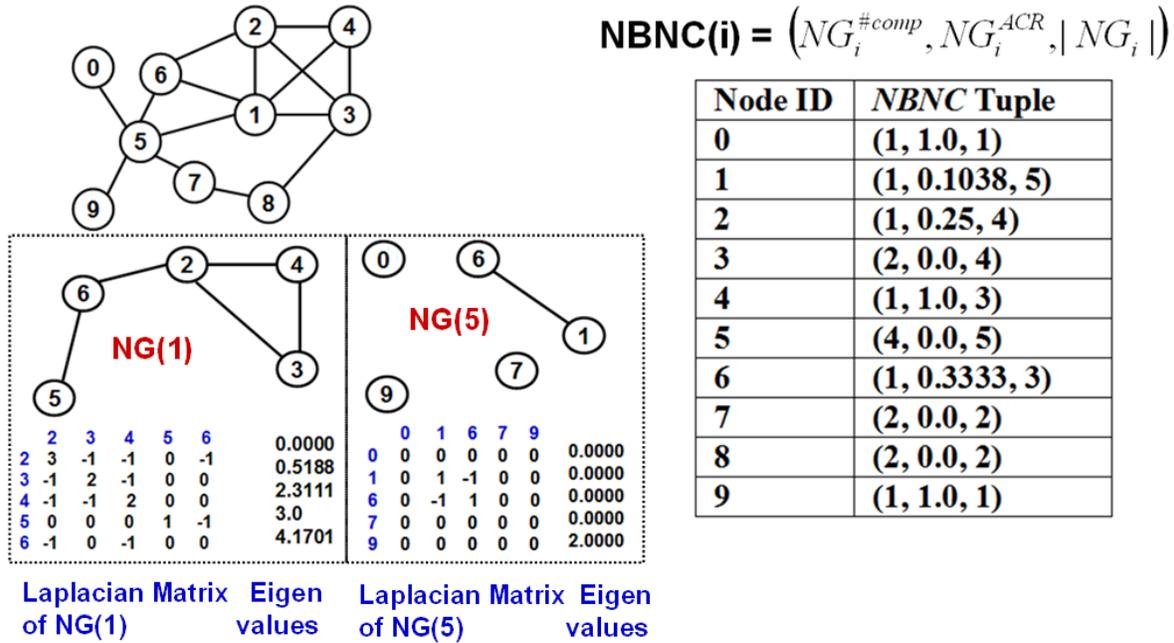

**Figure 1:** Example Graph and the NBNC Tuples of its Vertices

## 3 Procedure to Simulate the SIS Model

We use the SIS (Susceptible-Infected-Susceptible) model [17] to evaluate the effectiveness of NBNC tuple-based vaccination of nodes vis-a-vis the DEG centrality-based vaccination. Under this model, a node remains either in susceptible state or infected state. A susceptible node could get infected with a probability $\beta$ and an infected node could recover with a probability $\mu$. Once recovered, the node enters the susceptible state. We start with a graph of '$N$' nodes and decide to vaccinate a certain fraction ($\lambda$) of nodes. The nodes that constitute the $\lambda$ fraction of nodes to be vaccinated are chosen based on their ranking with respect to either the NBNC tuple or the DEG centrality values.

As part of initialization, we generate a random number for each node and set the node to either susceptible state (if the random number generated is greater than $\beta$) or infected state (if the random number generated is less than or equal to $\beta$). The simulation proceeds in rounds. Each round has two phases, executed in this sequence: (phase-i) Infected nodes changing their state to Susceptible with a probability $\mu$; (phase-ii) Nodes that still remain infected after phase-i will infect their susceptible neighbors (i.e., neighbor nodes that are neither vaccinated nor infected) with a probability $\beta$. We count the number of nodes that are in the Infected state after phase-ii and add it to the total number of infected nodes across all the rounds. The simulation is stopped in one of these two ways: (1) A round of simulation is run if at least one node stays Infected after phase-i of the round; otherwise, the simulation stops. (2) We run the simulations for a maximum number of rounds and then stop. After the simulation has stopped, we determine the average number of infected nodes per round by dividing the total number of infected nodes across all the rounds by the total number of rounds the simulation was run.

**Simulation of a Sample Round:** Figure 2 presents the execution of a sample round of the simulation. Let the parameters $\beta$ and $\mu$ be 0.5 each. Figure 2-(a) shows the graph at the beginning of the round (i.e., before phase-i is executed). Node 1 (in green color) is vaccinated; nodes 0, 3, 6, 7 and 9 are in the Infected state (pink in color); nodes 2, 4, 5 and 8 are in Susceptible state (yellow in color). To execute phase-i, we generate a random number (in the range of 0...1) at each infected node: if the random number comes out to be less than or equal to $\mu$, the infected node is considered to have recovered and moves from Infected state to Susceptible state. In Figures 2-(a) and 2-(b), we notice nodes 0, 6 and 9 (with a random number less than 0.5 for each of them) recover and become susceptible. At the end of phase-i (which is also before the beginning of phase-ii), nodes 0, 5, 6, 8 and 9 are susceptible; nodes 2, 3, 4, and 7 are infected. During phase-ii, for each infected node, we generate a random number for the link with its susceptible neighbors: if the random number comes out to be less than or equal to $\beta$, then that susceptible neighbor node is considered to have become infected. Both the neighbors of node 8 are infected; there is a 50% chance that node 8 could get infected due to one of them and it happens so due to node 7. On the other hand, among the three neighbors of node 6, only one of them (node 2) is in the Infected state (and the other two nodes are susceptible and vaccinated; so no infection could occur due to these two nodes); the chances of node 6 becoming infected is only 1/3 and the random number generated for the link 2-6 is greater than $\beta = 0.5$. Similarly, for node 5, only one of its five neighbors (node 7) are in the Infected state and it is less likely for the node to become infected in this round and the node does appear to remain Susceptible with the random number (0.7) generated for the link 5-7 being greater than $\mu = 0.5$.

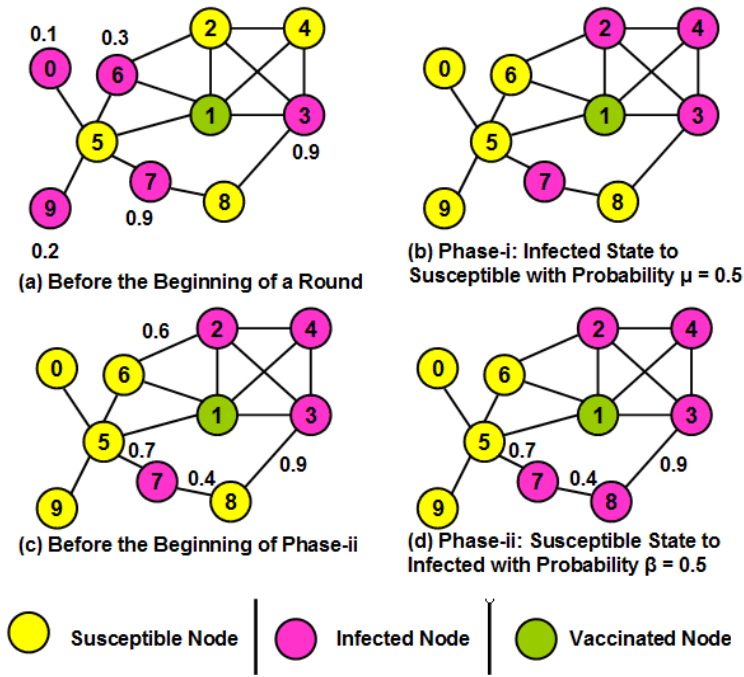

**Figure 2:** A Sample Round of Simulation: Phase-i and Phase-ii

A susceptible node that is surrounded only by infected neighbor nodes manages to stay in Susceptible state only if the random numbers generated for each of those links with the infected neighbor nodes is greater than $\beta$. This is where vaccination comes helpful. If a vaccinated node is the only neighbor for a susceptible node, then the latter will stay forever as susceptible and will never become infected. Even if a vaccinated node is one of the few neighbors of a susceptible node, the susceptible node is less likely to become infected during any round. On the other hand, if a susceptible node is surrounded by several infected neighbor nodes, the node is likely to become infected during any round.

**Simulation of Multiple Rounds (NBNC vs. DEG):** Figures 3 and 4 present the simulation of the SIS model for a total of five rounds with parameters $\beta$ and $\mu$ to be 0.5 each and 10% ($\lambda = 0.1$) of the nodes are to be vaccinated (in a graph of 10 nodes, $\lambda$ value of 0.1 corresponds to one node). Figure 3 demonstrates the infection spread in the presence of a NBNC-based vaccinated node (node 5 is the top ranked node on the basis of NBNC) and Figure 4 demonstrates the infection spread in the presence of DEG-based vaccinated node (node 1 is chosen over node 5 for vaccination, even though both have the larger DEG centrality). We follow the same color coding (yellow for susceptible nodes; green for vaccinated node and pink for infected node) as in Figure 2 for Figures 3 and 4 as well.

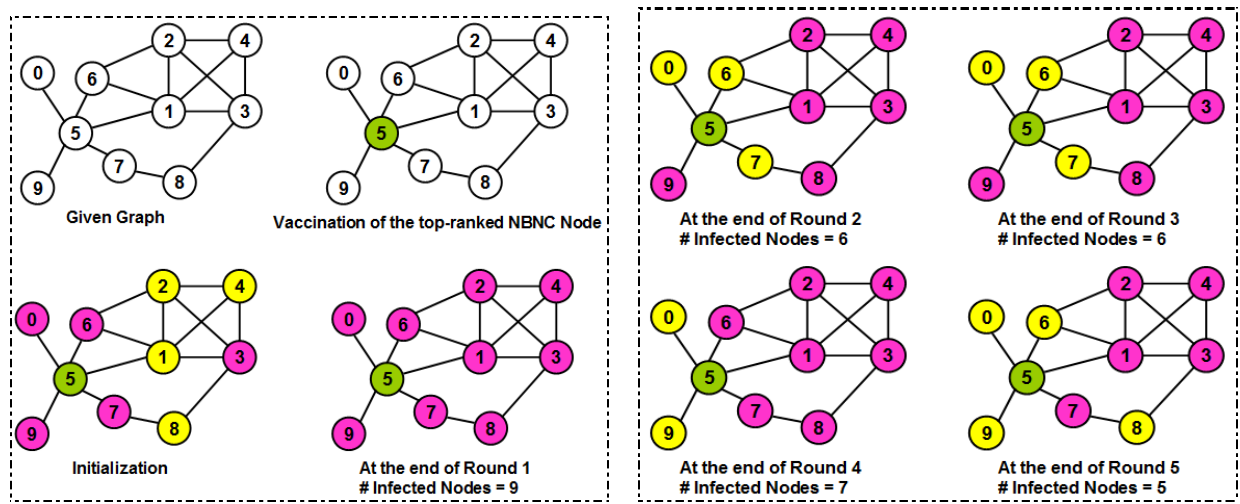

Figure 3: SIS Simulation with NBNC Tuple-based Selection of Nodes for Vaccination

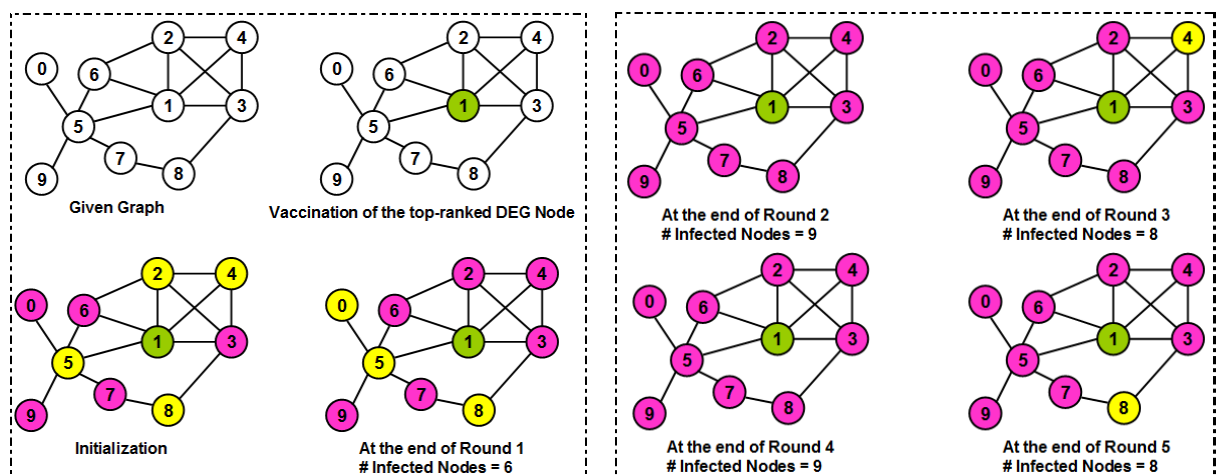

Figure 4: SIS Simulation with Degree Centrality (DEG)-based Selection of Nodes for Vaccination

**Figure 3:** The advantage with node 5 (the top-ranked NBNC node in the graph of Figure 3) is that its two neighbors nodes 0 and 9 have only one neighbor node (node 5) and that neighbor node is vaccinated. Hence, once nodes 0 and 9 recover from the infection and become susceptible, they will stay as susceptible and never get infected, because their only neighbor node (node 5) is a vaccinated node.

Likewise, one of the two neighbors of node 7 is a vaccinated node (node 5): hence, node 7 could get infected only if its other neighbor node (node 8) is infected and the random number generated for the link 7-8 is less than or equal to $β$. Overall, we notice that the presence of a vaccinated node connecting neighbor nodes (that would otherwise be disconnected or sparsely connected through multi-hop paths) provides herd immunity to the susceptible nodes and maintain them in the same state without getting infected. The average number of infected nodes per round with NBNC-based vaccination (Figure 3) is (9 + 6 + 6 + 7 + 5)/5 = 6.6. **Figure 4:** With node 1 as the vaccinated node, we still observe the infection to spread across the neighbors of node 1. Nodes 2 and 4 are initially susceptible and one of their other two neighbor nodes is vaccinated. Still, we observe node 2 and node 4 to be infected due to nodes 6 and 3 respectively. Because of node 6, node 5 gets infected as well. Due to node 5, nodes 0 and 9 get infected as well. We notice the infection spread to bypass node 1 and circulate among its neighbors as well as spread further to far away nodes. The average number of infected nodes per round with DEG-based vaccination (Figure 4) is (6 + 9 + 8 + 9 + 8)/5 = 8.0, which is much greater than 6.6, observed for NBNC-based vaccination. Similar results are observed for real-world networks as well (see Section 4).

## 4  SIS Simulations for Real-World Networks

We conducted simulations of the infection spread (per the SIS model) for a collection of 10 real-world networks, predominantly social networks. Table 1 lists the networks and their IDs used in Figures 5-6. The simulation parameters for the SIS model are: ($β$) - the probability with which a susceptible node becomes an infected node during any round; ($μ$) - the probability with which an infected node gets cured and returns to the Susceptible state during any round. We also vaccinate $λ$ fraction of nodes for any simulation and the nodes to be vaccinated are chosen based on their ranking with respect to either the NBNC tuple or DEG (degree centrality). The values of {$β$}, {$μ$} and {$λ$} used are: {0.3, 0.5, 0.7}, {0.25, 0.5} and {0.05, 0.10, 0.15, 0.20, 0.30} respectively. The operating conditions of a simulation on a particular real-world network are each possible combination of the above different values of the parameters $β$, $μ$ and $λ$ as well as the vaccination strategy (NBNC or DEG-based), leading to a total 3 * 2 * 5 * 2 = 60 operating conditions. We conduct 50 trials of the simulations for each operating condition and each simulation is run for a maximum of 20 rounds (a simulation stops prematurely before 20 rounds, if no nodes are infected for the next round) and we measure the average number of infected nodes per round (across all the rounds and trials) for each of the 10 real-world networks under each of the 60 operating conditions. We compute two metrics based on the average number of infected nodes per round for each network: (1) The average fraction of infected nodes per round for NBNC vs. DEG: computed as the ratio of the average number of infected nodes per round and the number of nodes for the network and (2) The ratio of the average fraction of infected nodes based on DEG as the node selection strategy for vaccination and the average fraction of infected nodes based on NBNC as the node selection strategy for vaccination.

Table 1: Real-world Networks used in the Simulations

| Net-ID | Network Name | # Nodes |
|---|---|---|
| Net-1 | Taro Exchange Network [20] | 22 |
| Net-2 | Sawmill Strikers Network [21] | 24 |
| Net-3 | Karate Network [22] | 34 |
| Net-4 | Teenage Women Friends Network [23] | 50 |
| Net-5 | Lazega Law Firm Network [24] | 71 |
| Net-6 | Copperfield Network [25] | 87 |
| Net-7 | US Football 2000 Network [26] | 105 |
| Net-8 | Anna Karnenina Network [27] | 138 |
| Net-9 | Jazz Band Network [28] | 198 |
| Net-10 | CKM Physicians Network [29] | 246 |

Figure 5 presents a heat map-based visualization of the numbers/results for the average fraction of infected nodes per round of the simulations for each of the 10 real-world networks and with respect to either NBNC or DEG as the node selection strategy for vaccination. The color coding in Figure 5 is based on the premise that lower the value for the average fraction of infected nodes per round, the more effective is the particular centrality-based node selection strategy for vaccination: accordingly, the colors of the cells featuring these raw values range from red (larger value for the average fraction of infected nodes per round) to green (lower value for the average fraction of infected nodes per round).

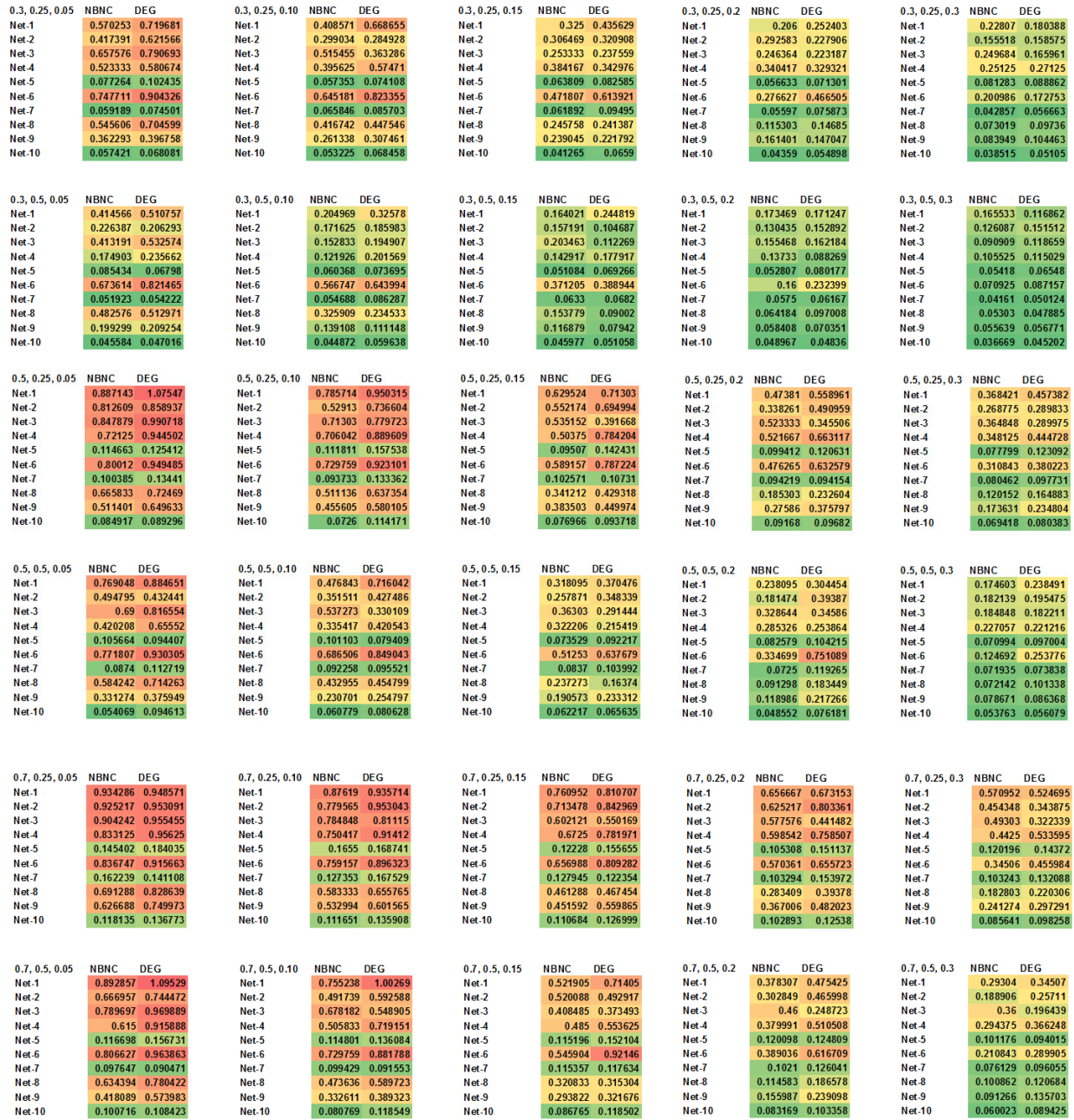

**Figure 5:** Heat Map Visualization of the Average Fraction of Infected Nodes Incurred per Round with respect to NBNC and DEG across all the Real-World Networks and Operating Conditions

As expected for both NBNC and DEG, the transition in the colors of the cells from red to green (in Figure 5) occurs gradually as we move across the scenarios exposing more nodes to infection (scenarios where $\beta \geq \mu$ and lower values of $\lambda$) to scenarios wherein nodes would not or are less likely to get infected (scenarios with larger values of $\lambda$ and $\beta < \mu$). Each row in Figure 5 corresponds to a particular combination of $\beta$ and $\mu$, with the $\lambda$ values increased from 0.05 to 0.30; the colors of the cells accordingly transition from red to green as we increase $\lambda$ for a given combination of $\beta$ and $\mu$. For most of the operating conditions and real-world networks, the average fraction of infected nodes with DEG is noticeably greater than that the average fraction of infected nodes with NBNC as the node selection strategy for vaccination. As a result, any particular record in Figure 5 (for a particular network, for a given combination of $\lambda$, $\beta$ and $\mu$) is more likely to have the cell for NBNC to be less red (or more yellow or more green) compared to the cell for DEG.

Figure 6 presents the ratio of the average number of infected nodes (DEG) and the average number of infected nodes (NBNC) for the 10 networks when the simulations are run for a particular combination of $\beta$ and $\mu$, with the $\lambda$ values increased from 0.05 to 0.30. We observe the ratio values to be heavily distributed above the line for 1.0, confirming our claim that NBNC would be more appropriate to choose nodes for vaccination vis-a-vis DEG. We observe fewer ratios to be less than 1.0 for scenarios in which more nodes are exposed for infection (i.e., when $\beta \geq \mu$). The median of the ratio values are typically in the range of 1.15 to 1.20 for most of the operating conditions.

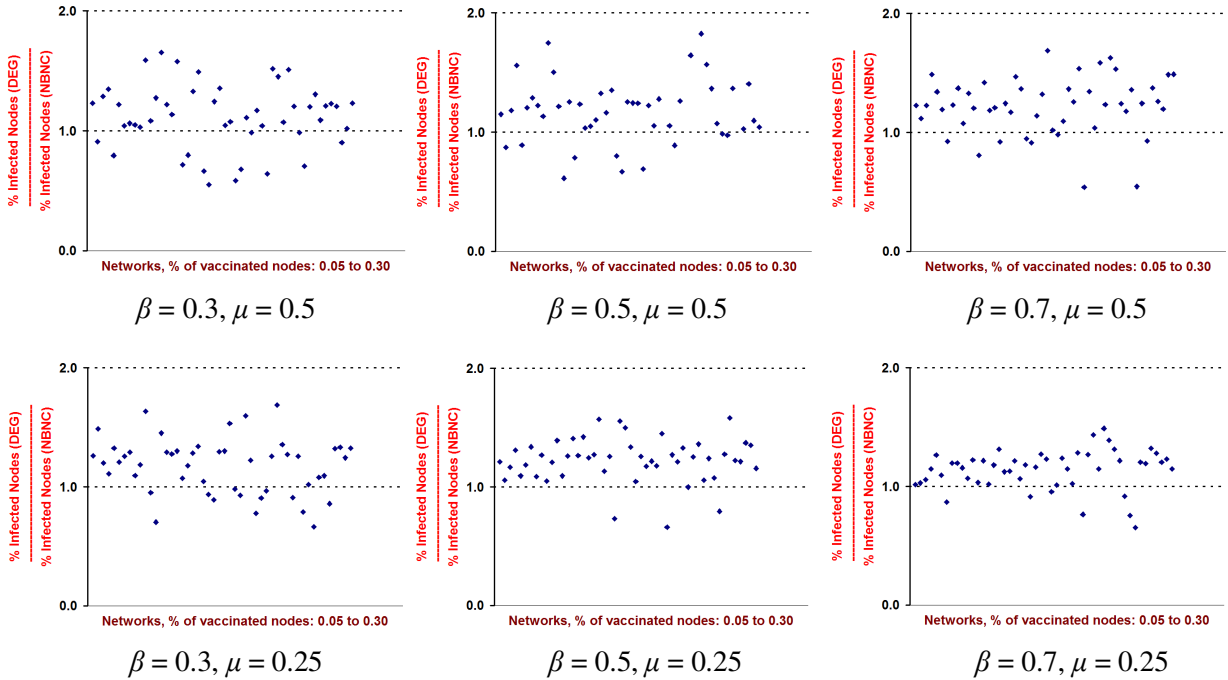

**Figure 6:** Ratio of the % Infected Nodes for the 10 Real-World Networks and for different Values of the Fraction of Nodes that are Vaccinated with DEG and NBNC as Node Selection Strategies for Vaccination

## 5 Related Work

In a recent work [1], centrality metrics were considered for vaccination to contain the spread of an infection per the SIR model. Unlike our work, in [1], the nodes chosen for vaccination and the nodes that recover from the infection are simply removed from the network; such a simulation approach will not work for the SIS model as (with the SIS model) the recovering nodes immediately become susceptible and need to be considered part of the network as a potential candidate for future infections. In [2], the authors observed that betweenness centrality [3] is effective in containing the spread of an infection in synthetic networks generated per the Barabasi-Albert scale-free model [4]; whereas, the degree centrality was found to be more effective in reducing the spread of an infection in real-world networks. Note that betweenness centrality is a computationally-heavy metric and its computation needs to be synchronous (i.e., the algorithm needs to be run at all the nodes in the network even if the metric is to be determined for a particular node) as well as requires global knowledge. Recently [9, 10], community-aware approaches have been evaluated for preferential vaccination of nodes; but, such strategies require the knowledge of global information all the time. On the other hand, the NBNC tuple [11] for a node can be computed using local neighborhood information of the node and its neighbors. Non-centrality and/or non-community-based strategies for vaccinating nodes have been typically found to be less effective [12] (for example: the strategy [13] of choosing the random neighbor of a randomly chosen node).

Per [5], the problem of identifying critical nodes for vaccination and protection of susceptible non-vaccinated nodes has been shown to be equivalent to the problem of identifying the super spreader nodes for information diffusion [6]. In [7], the authors showed that given an adjacency matrix-style contact matrix of people in a social network (without any constraint on the structure of the matrix), the diffusion probability of the information has been observed to be proportional to the largest Eigenvalue (a.k.a. the spectral radius [8]) of the adjacency matrix; the lower the spectral radius, the lower the diffusion probability of the information (also applicable for viral diffusion).

## 6 Conclusions and Future Work

The high-level contribution of this paper is a simulation-based approach to identify the most effective centrality-based node selection strategy for preferential vaccination of nodes to reduce the fraction of infected nodes during any spread of an infection. We used the SIS model for the infection spread. The simulation and evaluation procedures employed in this paper can be used for any measure adopted for preferential vaccination of nodes. We observe the neighborhood-based bridge node centrality (NBNC) approach to be more effective compared to the widely considered degree centrality (DEG) approach for preferential vaccination. We thus show that the bridge node-based NBNC tuple is more effective in identifying node(s) whose neighbors are less likely to infect each other if the node(s) are vaccinated. This is the key factor behind the success of NBNC over DEG towards reducing the number of infected nodes per round, especially with increase in the fraction of vaccinated nodes. As part of future work, we plan to evaluate the effectiveness of NBNC vs. DEG for preferential node vaccination with respect to the familiar infection spread models such as the SIR and SIRS models [17] involving both real-world networks as well as synthetic networks generated from the theoretical scale-free network (e.g., [4]) and small-world network models (e.g., [30]).

**Acknowledgment**
The work leading to this paper was partly funded through the U.S. National Science Foundation (NSF) grant OAC-1835439 and partly supported through a sub contract received from University of Virginia titled: Global Pervasive Computational Epidemiology, with the National Science Foundation as the primary funding agency. The views and conclusions contained in this paper are those of the authors and do not represent the official policies, either expressed or implied, of the funding agency.

**References**
[1] F. Sartori, M. Turchetto, M. Bellingeri, F. Scotognella, R. Alfieri, N.-K. -K. Nguyen, T.-T. Le, Q. Nguyen and D. Cassi, "A Comparison of Node Vaccination Strategies to Halt SIR Epidemic Spreading in Real-World Complex Networks," *Scientific Reports*, vol. 12, no. 21355, pp. 1-13, 2022.
[2] X. Wei, J. Zhao, S. Liu and Y. Wang, "Identifying Influential Spreaders in Complex Networks for Disease Spread and Control," *Scientific Reports*, vol. 12, no. 5550, pp. 1-11, 2022.
[3] U. Brandes, "A Faster Algorithm for Betweenness Centrality," *The Journal of Mathematical Sociology*, vol. 25, no. 2, pp. 163-177, 2001.
[4] A-L. Barabasi and R. Albert, "Emergence of Scaling in Random Networks," *Science*, vol. 286, no. 5439, pp. 509-512, 1999.

[5] R. Paluch, X. Lu, K. Suchecki, B. K. Szymański, J. A. Hołyst, "Fast and Accurate Detection of Spread Source in Large Complex Networks," *Scientific Reports*, vol. 8, no. 1, pp. 1-10, 2018.
[6] D. Zhang, Y. Wang and Z. Zhang, "Identifying and Quantifying Potential Super-Spreaders in Social Networks," *Scientific Reports*, vol. 9, no. 14811, pp. 1-11, 2019.
[7] Y. Wang, D. Chakrabarti, C. Wang and C. Faloutsos, "Epidemic Spreading in Real Networks: An Eigenvalue Viewpoint," *Proceedings of the 22nd International Symposium on Reliable Distributed Systems*, pp. 25–34, 2003.
[8] J-M. Guo, Z-W. Wang, X. Li, "Sharp Upper Bounds of the Spectral Radius of a Graph," *Discrete Mathematics*, vol. 342, no. 9, pp. 2559-2563, 2019.
[9] H. Cherifi, G. Palla, B. K. Szymanski and X. Lu, "On Community Structure in Complex Networks: Challenges and Opportunities," *Applied Network Science*, vol. 4, no. 117, pp. 1-35, 2019.
[10] Z. Ghalmane, C. Cherifi, H. Cherifi and M. El Hassouni, "Centrality in Complex Networks with Overlapping Community Structure," *Scientific Reports*, vol. 9, no. 10133, pp. 1-29, 2019.
[11] N. Meghanathan, "Neighborhood-based Bridge Node Centrality Tuple for Complex Network Analysis," *Applied Network Science*, vol. 6, no. 47, pp. 1-36, 2021.
[12] T. Lev and E. Shmueli, "State-based Targeted Vaccination," *Applied Network Science*, vol. 6, no. 6, pp. 1-16, 2021.
[13] L. K. Gallos, F. Lilijeros, P. Argyrakis, A. Bunde and S. Havlin, "Improving Immunization Strategies," *Physical Review E*, vol. 75, no. 045104, April 2007.
[14] M. E. J. Newman, *Networks: An Introduction*¸ Oxford University Press, Oxford, UK, 1st Edition, September 2010.
[15] J. Ma, P. van den Driessche and F. H. Willeboordse, "The Importance of Contact Network Topology for the Success of Vaccination Strategies," *Journal of Theoretical Biology*, vol. 325, pp. 12-21, May 2013.
[16] K. Musial and K. Juszczyszyn, "Properties of Bridge Nodes in Social Networks," *Proceedings of the International Conference on Computational Collective Intelligence: Lecture Notes in Artificial Intelligence*, vol. 5796, pp. 357-364, 2009.
[17] J. Liu and S. Xia, *Computational Epidemiology: From Disease Transmission Modeling to Vaccination Decision Making*, Springer, 1st edition, September 2020.
[18] M. Fiedler, "Algebraic Connectivity of Graphs," *Czechoslovak Mathematical Journal*, vol. 23, no. 98, pp. 298-305, 1973.
[19] G. Strang, G. *Linear Algebra and its Applications*, 4th Edition, Brooks Cole, Pacific Grove, CA, USA, 2006.
[20] E. Schwimmer, *Exchange in the Social Structure of the Orokaiva: Traditional and Emergent Ideologies in the Northern District of Papua*, C Hurst and Co-Publishers Ltd., December 1973.
[21] J. H. Michael, "Labor Dispute Reconciliation in a Forest Products Manufacturing Facility," *Forest Products Journal*, vol. 47, no. 11-12, pp. 41-45, October 1997.
[22] W. W. Zachary, "An Information Flow Model for Conflict and Fission in Small Groups," *Journal of Anthropological Research*, vol. 33, no. 4, pp. 452-473, 1977.
[23] M. Pearson and L. Michell, "Smoke Rings: Social Network Analysis of Friendship Groups, Smoking and Drug-taking," *Drugs: Education, Prevention and Policy*, vol. 7, no. 1, pp. 21-37, 2000.
[24] E. Lazega, *The Collegial Phenomenon: The Social Mechanisms of Cooperation Among Peers in a Corporate Law Partnership*, Oxford University Press, 1st Edition, September 2001.
[25] M. E. J. Newman, "Finding Community Structure in Networks using the Eigenvectors of Matrices," *Physical Review E*, vol. 74, no. 3, 036104, September 2006.
[26] M. Girvan and M. E. J. Newman, "Community Structure in Social and Biological Networks," *Proceedings of the National Academy of Sciences of the United States of America*, vol. 99, no. 12, pp. 7821-7826, June 2002.
[27] D. E. Knuth, *The Stanford GraphBase: A Platform for Combinatorial Computing*, 1st Edition, Addison-Wesley, Reading, MA, December 1993.

[28] P. Geiser and L. Danon, "Community Structure in Jazz," *Advances in Complex* Systems, vo. 6, no. 4, pp. 563-573, July 2003.

[29] C. Van den Bulte and G. L. Lilien, "Medical Innovation Revisited: Social Contagion versus Marketing Effort," *American Journal of Sociology*, vol. 106, no. 5, pp. 1409-1435, March 2001.

[30] D. J. Watts and S. H. Strogatz, "Collective Dynamics of Small-World Elements," *Nature*, vol. 393, pp. 440-442, June 1998.